# Errors of LLM-Assisted Literature Retrieval in Environmental Science: A Comparison Study of Abstract versus Full-text Based Prompts

Yanjun Chen,[1,2] Yongfeng Zhang,[3] Lanjing Zhang[1, 3, 4, 5, 6 *]

[1]Department of Chemical Biology, Ernest Mario School of Pharmacy, Rutgers University, Piscataway, NJ, USA.

[2] East Brunswick High School, East Brunswick, NJ, USA.

3 Department of Computer Science, Rutgers University, Piscataway, NJ, USA.

[4] Department of Pathology, Princeton Medical Center, Plainsboro, NJ, USA.

[5] Department of Pharmacology, Physiology, and Neuroscience, New Jersey Medical School, Rutgers University, Newark, NJ, USA.

[6] Rutgers Cancer Institute, New Brunswick, NJ, USA.

*Correspondence to: Lanjing Zhang, Department of Chemical Biology, Ernest Mario School of Pharmacy, Rutgers University, Piscataway, NJ, 08854 USA. Phone (848) 445-8004, Fax (732) 445-0687, lanjing.zhang@rutgers.edu

**Key words:** Large language model, Literature retrieval, Research methods, Environment science, Information Retrieval

# Abstract (250 words):

Large language models (LLMs) are increasingly used for literature search and synthesis. However, it is unclear whether they retrieve accurate bibliographic information in environmental science. Therefore, we quantitatively compared the errors of widely used LLM platforms in retrieving references related to original articles from five leading environmental science journals (*Energy and Environmental Science, Nature Sustainability*, *Nature Climate Change, Lancet Planetary Health,* and *Environmental Science and Technology*) published in 2024–2025. Claude, ChatGPT, Grok, DeepSeek, Perplexity, and Gemini were used as the LLM platforms. LLMs retrieved 10 references for each of the 50 randomly selected original article using either the article's abstract or its full-text as prompt. The retrieved references were subject to a multimetric score ratio combining validity of bibliographic data, Google Scholar link, digital object identifier, Scopus Electronic Identifier and relevance score (cited by or being the index paper), and the proportion of complete fabrication that failed all metrics. Abstract-only prompt yielded significantly higher accuracy than full-text one. This advantage was confirmed in multilevel mixed-effect multivariable regression after adjusting for journal, platform, and output order. Source journal and the position of a reference within the output list were also independently associated with retrieval accuracy, with lower-listed references associated with lower accuracy. These findings suggest that LLM-assisted literature retrieval in environmental science remains moderately accurate and overall inconsistent, varying significantly by platform, journal, prompt type, and output position. Abstract-based prompting, as task-aligned information compression, may outperform full-text one in literature retrieval. Caution should be used when generalizing our findings.

# 1. Introduction

Large Language models (LLMs) are rapidly used in scientific writing and literature review, including literature search, data extraction, evidence synthesis, and drafting manuscript [1-4]. Indeed, LLM-assisted writing in engineering increases citations in a study of more than 1 million articles [5]. In environmental science, LLMs have also been explored for ecological information extraction, retrieval of environmental impact data from scientific literature, environmental decision making and research opportunity identification [6-9]. The involved areas of environmental science include hydrology research [10], plastic recycling [11], and environmental sustainability [12].

One of the major challenges of LLM-assisted literature analysis is fabricated articles and incorrect bibliographic metadata [13-15]. The reported precision of retrieving a valid title ranged from 16.6% to 64.8%, while that of retrieving valid first-author name ranged from 7.0% to 55.8% [16]. Similar findings were documented in medicine, bioinformatics and natural language processing [14, 16-18]. Efforts have thus been made to reduce these errors [1, 19], such as retrieval-augmented generation (RAG) [1, 3, 19, 20].

While others mostly focused on the accuracy of scientific data or research trends in environmental science [7-9], similar concerns over retrieval errors have also been raised in ecology and sustainability [12, 21]. However, these concerns are mostly theoretical and anecdotal. There is also little data on the errors in LLM-assisted literature retrieval in environmental science. It remains unclear how to best reduce these errors. Indeed, there are no data on the error rate in retrieving references of environmental science as shown by recent systematic and narrative reviews and one position paper from the Environmental Data Science Summit [22-24], although they all advocate responsible use of LLMs.

Past studies appear to support prompt compression. First, prompt compression can discard non-task-relevant information and overcome information bottle, yet retaining the information predictive of or associated with the targets [25]. Second, the length of prompt and information location in the prompt strongly influences LLM performances, with lower performances in the longer ones [26]. Finally, Jiang at el. have shown that prompt compression could not only reduce computing costs and time but improve performance by up to 21% [27, 28]. Xu et al. demonstrated that selecting sentences or generating summary can effectively reduce prompt length while maintaining or improving LLM’s performance [29]. However, few studies have tested whether human-generated semantic compression, such as abstract of scientific publications, improve LLM performances.

The theoretical framework also seems to support prompt compression and intrigued us to compare abstract versus full-text based prompting in this study. Retrieval accuracy can be formulated as trading off two terms that move in opposite directions as one goes from full text to abstract: A. Noise-interference cost that falls with compression. There are fewer irrelevant tokens competing for attention, less risk of the answer landing in a "lost-in-the-middle" position, and higher density of task-relevant tokens per context window. B. Missing-evidence cost that rises with compression: These are the claims, caveats, quantitative detail, and supporting evidence that exist only in the full text and are never fully captured by the abstract's human-compression process. The scientific abstracts are favored as the prompt because they are the distilled products of authors, editors and reviewers. Their scientific insights and information are not only intense (compressed) but highly relevant. Thus, abstract-based prompt may have the advantages of being human-generated semantic compression.

Therefore, we conducted this quantitative comparison study to understand the errors in LLM-assisted literature retrieval in environmental science and their associated factors, focusing on abstract versus full-text based prompts. Specifically, we analyzed the errors of six widely used free-tier LLM platforms in retrieving papers relevant to original research articles published in three leading environmental science journals. We also sought the factors independently associated with these retrieval errors, focusing on the input method of using abstract versus full-text prompt.

# 2. Materials and Methods

### 2.1 Random selection of original articles

Five original research articles published in 2024 or 2025 were randomly selected from each of the three leading environmental science related journals, namely *Energy and Environmental Science, Nature Sustainability, Nature Climate Change, Lancet Planetary Health,* and *Environmental Science and Technology*. We selected all original research articles from these three journals published in 2024 and 2025 from Scopus, categorized them by year, and assigned them each a unique number. We then used Google's random number generator to generate five random numbers for each journal for both 2024 and 2025. Only articles corresponding to the randomly generated numbers were included [14].

### 2.2 LLM assisted literature retrieval

We conducted separate searches using the abstracts or full-texts of the 50 original articles as described before [14]. Six commonly used LLM platforms (free versions) were chosen and used, including Claude (Sonnet 4.6), ChatGPT (GPT-5.3), Grok (Grok 4 or Grok 4.2), DeepSeek (DeepSeek V3), Perplexity (Nemotron 3 Super), and Gemini (Gemini 3.5 Flash). It is noteworthy that we subscribed to the LLMs to use their legacy (then free-tier) models when collecting additional data in August 2026.

For each search, we provided two prompts to each LLM platform along with the abstracts or full-texts of the 30 selected original articles. The first prompt was "Generate 10 key peer-reviewed journal references related to this content in environmental science. For each reference, provide title, first author, journal, year, the volume, the issue, the first page number, google scholar link, EID and DOI if available" with the article's abstract presented at the end of the prompt. We then prompted the LLM platforms to confirm these publication bibliographic data using "Confirm the title, first author, journal, year, the volume, the issue, the first page number, google scholar link, EID and DOI for each reference." The results confirmed by the platform were recorded and subject to manual checking.

For literature retrieval based on full texts, we first copied the full text into a blank document, removed references, and then submit that document to the LLM platforms. References were removed to eliminate the possible referral biases.

**2.3 Manual validation of retrieved references**

We first verified each retrieved reference using multiple bibliographic identifiers and reference information. We searched and checked the digital object identifier (DOI), google scholar link, Scopus Electronic Identifier (EID), first author's name, and other bibliographic information. If the information provided by the LLM matched the verified record, one point would be given and assigned to the corresponding metric. If the information was present but did not match the verified record, zero point would be assigned. If the information was resulted but did not match the verified record, zero point would be assigned.

We then assessed the relevance by comparing each retrieved reference with the reference list of the corresponding index article. If the retrieved reference was the index article or cited by the index article, one point would be assigned for relevance. Otherwise, zero point would be assigned.

**2.4 Calculation score ratio and complete fabrication status**

First, we generated the total score by adding six scores of each retrieved reference: relevance score, author score, bibliography score, Google Scholar link score, DOI score, and EID score. Because no not-applicable category was used in the current dataset, so all six metrics were included for every retrieved reference. Therefore, the score cap was fixed at six. The score ratio was calculated by dividing the total score by the score cap.

To identify references without any validated information we used the complete fabrication status as the metric. Specifically, a retrieved reference would be classified as a complete fabrication if its total score was zero.

### 2.5 Regression analyses and statistics

Descriptive statistics were used to summarize score ratio, complete fabrication rate and component-level accuracy across input type, LLM platform, publication year, source journal, and LLM output order. Pairwise comparisons were conducted using Mann-Whitney U tests for independent-group comparisons[30, 31]. Pairwise p-value matrices were generated for comparisons among LLM platforms and source journals.

Multilevel mixed-effects linear and logistic regression models (mixed /mlogit in Stata) with random intercepts for higher-level clusters (articles, and LLM platforms) were used to identify factors associated with score ratio and complete fabrication status, respectively. Multilevel mixed-effects models were used to account for the hierarchical structure of our data, while multivariable linear regression and multivariable logistic regression were also used for the sensitivity analysis. Predictors included input type, publication year, source journal, LLM platform, and output sequence. All statistical analyses and plotting were performed using python (version 3.12) and Stata (version 18. College Station, TX: StataCorp LLC.). Statistical significance was defined as two-sided p value $< 0.05$.

## 3. Result

### 3.1 Random selection of the original articles

The five leading environmental science journals published 2,979 original articles from January 1 to December 31, 2024 and 3,129 original articles from January 1 to December 31, 2025. Five original articles per year were randomly selected from these articles (**Figure 1**).

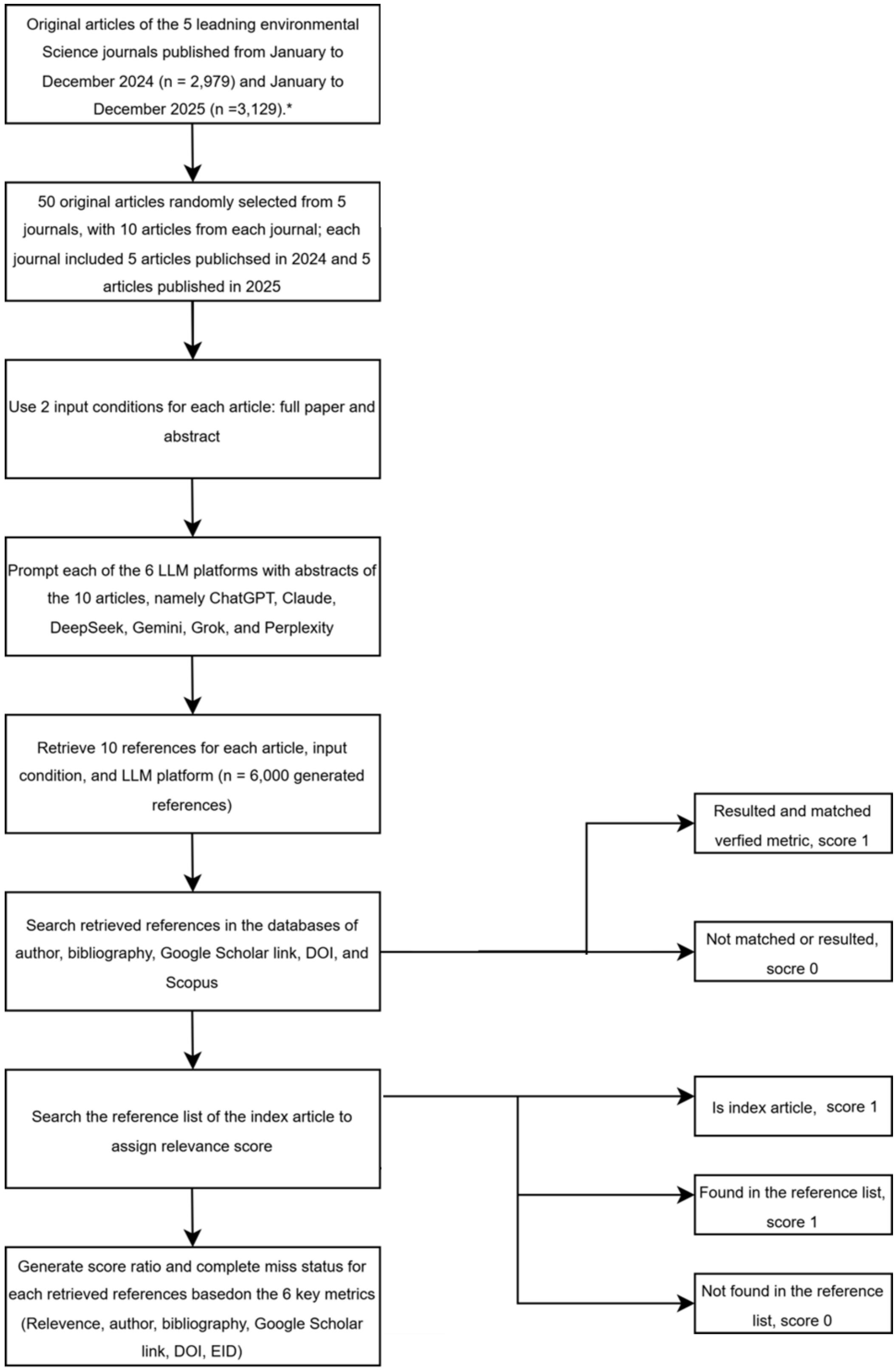
Original articles of the 5 leadning environmental Science journals published from January to December 2024 (n = 2,979) and January to December 2025 (n =3,129).*
50 original articles randomly selected from 5 journals, with 10 articles from each journal; each journal included 5 articles publichsed in 2024 and 5 articles published in 2025
Use 2 input conditions for each article: full paper and abstract
Prompt each of the 6 LLM platforms with abstracts of the 10 articles, namely ChatGPT, Claude, DeepSeek, Gemini, Grok, and Perplexity
Retrieve 10 references for each article, input condition, and LLM platform (n = 6,000 generated references)
Search retrieved references in the databases of author, bibliography, Google Scholar link, DOI, and Scopus
Resulted and matched verfied metric, score 1
Not matched or resulted, score 0
Search the reference list of the index article to assign relevance score
Is index article, score 1
Found in the reference list, score 1
Not found in the reference list, score 0
Generate score ratio and complete miss status for each retrieved references basedon the 6 key metrics (Relevence, author, bibliography, Google Scholar link, DOI, EID)

**Figure 1.** Flowchart of the study.

Note: The five leading environmental science journals including *Energy and Environmental Science, Nature Sustainability, Nature Climate Change, Lancet Planetary Health,* and *Environmental Science* and *Technology*; The six key accuracy metrics were bibliography, digital object identifier, relevance, Google Scholar link, author, Scopus Electronic Identifier. DOI, digital object identifier; EID, Scopus Electronic Identifier; LLM, Large Language Model

### 3.2 Overall score ratio and complete fabrication

Across both full-text (n=500) and abstract-only (n=500) input methods, 6,000 LLM-retrieved references were analyzed across the six LLM platforms. The mean score ratio was 0.42 (standard deviation 0.32; range, 0-1). A higher score ratio indicates greater accuracy in retrieving valid bibliographic information from environmental science literature.

### 3.3 Full-text input compared to Abstract-only input

Compared with full-text input, abstract-only input yielded a higher mean score ratio (0.43 versus 0.40, $p < 0.001$). The complete fabrication rate was lower for abstract-only input than for full-text input (27.8% versus 30.4%, $p = 0.027$). Detailed analyses showed that abstract-only input was linked to higher scores in DOI, EID, complete fabrication and score ratio, but not scores of relevance or Google Scholar link (**Table 1**).

### 3.4 Performance differences across LLM platforms

Score ratio varied significantly across LLM platforms (**Figure 2**). Claude had the highest mean score ratio (0.64), followed by Grok (0.60), ChatGPT (0.40), DeepSeek (0.39), Perplexity (0.32), and Gemini (0.14, **Supplementary Table 1**). Claude also had the lowest complete fabrication rate (4%), whereas Gemini had the highest complete fabrication rate (71.0%).

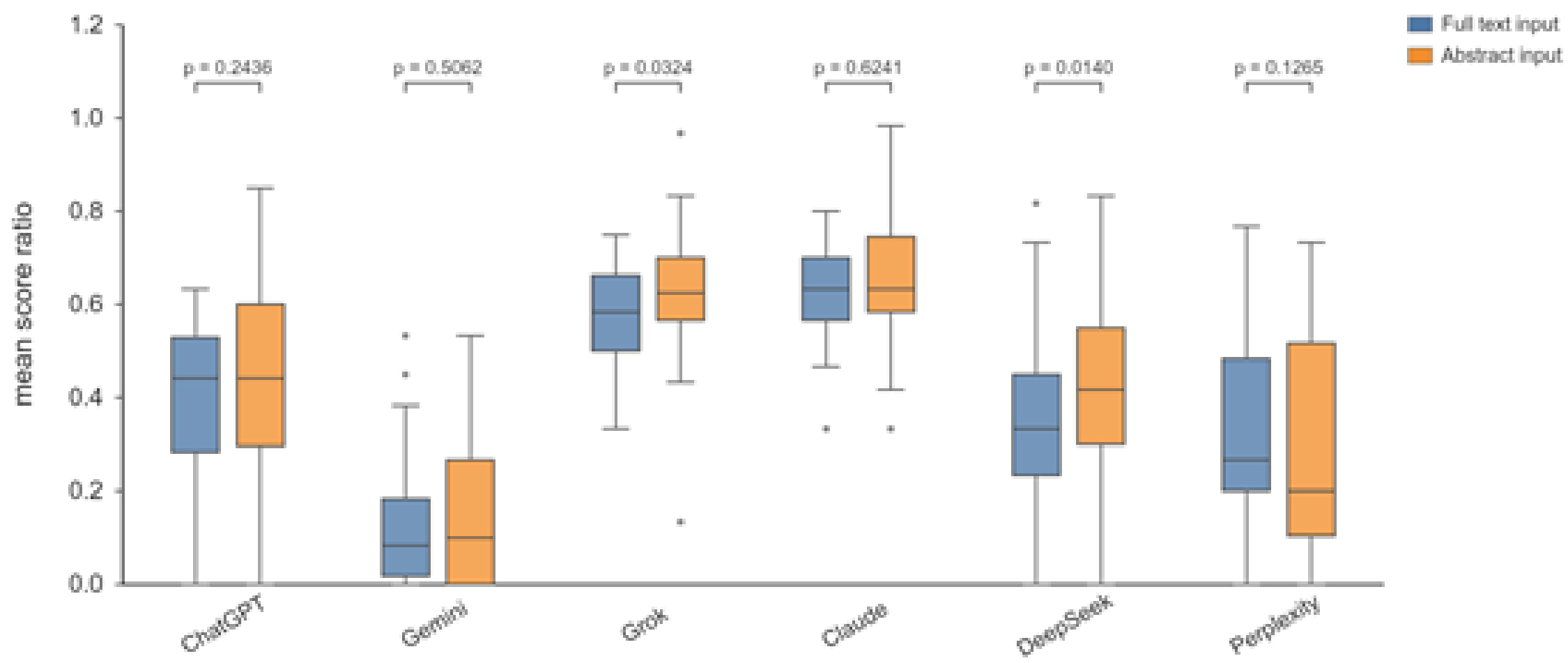

**Figure 2.** Mean Score Ratio by LLM Platforms and Input Type

Note: For each original article, LLM platform, and input type, the score ratios of the 10 retrieved references were averaged to generate one paper-level mean score ratio. Therefore, each boxplot contains 50 paper-level mean.

When full-text and abstract-only inputs were compared within each LLM platforms (**Figure 2**), abstract-only input was linked to a higher score ratio for Grok (p = 0.03) and DeepSeek (p = 0.001). In contrast, no statistically significant difference in full-text versus abstract-only was observed for the other LLMs.

### 3.5 Rates of retrieving valid bibliographic metrics across LLM Platforms and input method

Component-level retrieval accuracy varied significantly across validation metrics and LLM platforms**.** The highest accuracies in ranking order were those in year (68.5%), source title (66.7%), volume (63.1%), author (60.5%), DOI (58.2%) and Google Scholar link (42.3%), while relevance score (25.4) and EID (12.3%) showed the lowest. Claude showed the most consistently high accuracy across most bibliographic components with full-text or abstract-only input. In contrast, Gemini and Perplexity showed lower component-level accuracy across most metrics. Abstract-only input generally showed similar or higher accuracy than full-text input for most components, particularly for EID.

### 3.6 Multivariable regression analysis of the factors associated with retrieval accuracy

Year was not linked to score ratio or complete fabrication status, and was thus not included in multivariable regression analyses. Multilevel mixed-effects multivariable regression showed that journal, input method, output sequence and LLM platform were independently associated with score ratio (**Table 2**). Our sensitivity analysis using multivariable linear regression model showed similar findings (**Supplementary Table 2**).

Inputting full-text was associated with lower score ratios than inputting abstract (adjusted coefficient, -0.03; 95% CI, -0.04 to -0.02; $p<0.001$). Claude and Grok were linked to higher score ratios than ChatGPT ($p<0.001$ for both), but Gemini and Perplexity linked to lower score ratios ($p<0.001$ for both). Compared with *Energy and Environmental Science*, the other journals had no association with the score ratio. The output sequence number of LLM retrieved references was associated with a lower score ratio (adjusted coefficient, -0.03; 95% CI, -0.03 to -0.03; $p<0.001$). In other words, the reference at a lower position of the retrieved-reference list/output likely had a lower score ratio.

Multilevel mixed-effects multivariable logistic regression also showed similar independent associations of journal, input method, output sequence and LLM platform with complete fabrication status (**Table 3**). Given the inverse relationship of score ratio and complete fabrication, the adjusted ORs were inversed. For example, full-text was linked to a higher rate of complete fabrication than abstract (adjusted OR, 1.20; 95% CI, 1.05 - 1.37; $p=0.008$), and a lower score ratio. Similar findings were shown in our sensitivity analysis using multivariable logistic regression model (**Supplementary Table 3**).

The adjusted associations of LLM with score ratio among LLM platforms differed significantly in most of the one-on-one comparisons (**Figure 3**). With abstract input, all pairwise comparisons were statistically significant except Grok versus ChatGPT ($p = 0.72$) and DeepSeek versus Gemini ($p = 0.15$). With full-text input, all pairwise comparisons were also significant, except Perplexity versus Grok ($p = 0.79$).

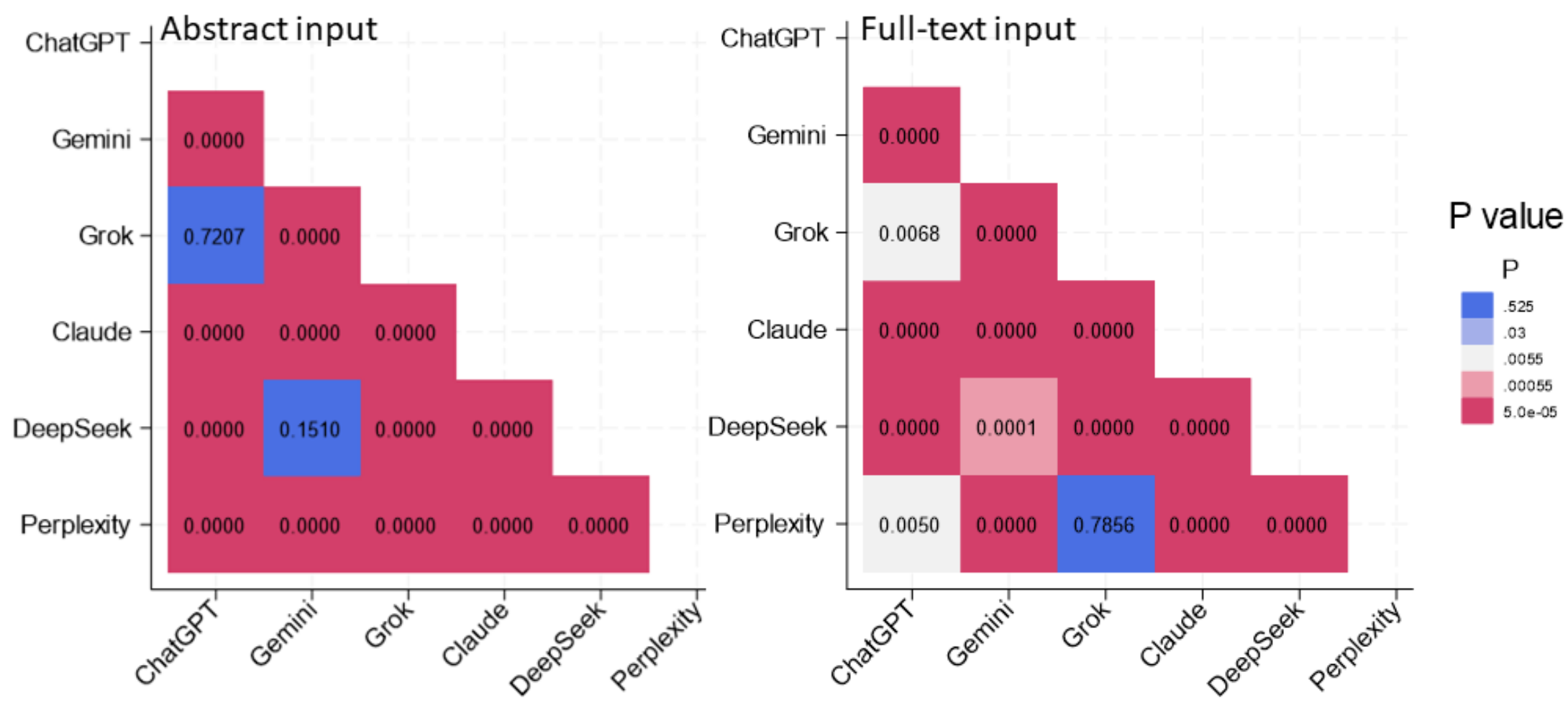

**Figure 3.** P-values of comparing association of LLM with score ratios between LLM platforms. Note: Left and right panels show abstract and full-text as the input method, respectively. The associations were calculated using multivariable mixed regression with journal and retrieved-reference's sequence as covariates.

### 3.7 Individual validity metrics

Across the 1,000 references retrieved by each LLM platform, performance varied across individual reference information fields by source journal or publication year (**Supplementary Table 4**). Significant platform-level differences were observed by source journal for EID, Google Scholar link, author, and bibliography, and by publication year for DOI, EID, Google Scholar link, author, and bibliography. DOI accuracy and Relevance score differed significantly by source journal ($p < 0.001$ for both), respectively, but not by publication year.

Claude and Grok generally retrieved accurate reference information across most fields more frequently than the average, but Gemini and Perplexity less frequently. For example, Claude showed consistently high counts for DOI, author, and bibliography, whereas Gemini had low counts for EID and Google Scholar link.

# 4. Discussion

### 4.1 Key implications

The key implications and novelties of this study are providing the first evidence of a small study on the errors in LLM-assisted literature retrieval in environmental science, and identifying the factors independently associated with these errors.

LLM-assisted literature retrieval seems to yield higher score ratios and fewer complete-fabricated references than reported in medical field [14]. Caution should be used when interpreting this data. Because the two studies used slightly different bibliographic metrics and the LLM platforms have evolved and improved rapidly. However, the moderate score ratio and complete fabrication rate as shown here are still persistent and concerning. They support the need of validating bibliographic information of the LLM-retrieved references in environmental science. They also warrant future studies to reduce the errors.

Moreover, we demonstrate the better performance of abstract-only over full-text inputting in retrieval accurate and relevant bibliographic data. Due to the shorter length of the abstract than full-text, it also saves computing time and resources. Thus, abstract inputting should be a choice over full-text in this regard.

This study cannot explain the reasons, but several possibilities and implications should be considered. First, prompt compression or shorter prompt has been shown to improve LLM performance [32-34] and retrieve more accurate bibliographic data in this study. Second, content quality seems to influence the retrieval performance. This hypothesis is supported by the journal's association with errors in LLM-assisted literature retrieval in this and other studies [14]. In other words, high-quality prompt/input such as abstract from high quality journals may help retrieve more accurate and relevant references. Finally, it appears necessary to understand whether some components of the abstract are more important than others for the accuracy in LLM-assisted literature retrieval. Maybe, an iterative process should be used to improve the accuracy as proposed before [21].

Furthermore, the newly-discovered association of output sequence with reference validity also has practical implications. References listed/ranked lower in the LLM output had lower score ratios and higher odds of complete fabrication as our data show. Therefore, they should be subject to more scrutiny than the ones listed/ranked higher in the output list. The finding also suggests that LLM models rank the retrieved references before generating the list. It will be important to understand the underlying algorithms and improve them.

Finally, we confirm the significant variability in errors of literature retrieval across LLM platforms and journals, as shown in medicine [14]. Our study extends this finding to environmental science. Careful selection of a high-performing LLM is thus recommended.

**4.2 Limitations.**

This study has limitations. First, we only used free versions of LLM platforms and evaluated with only two prompts. The advantages of wide-availability and unparalleled affordability of free-tier versions seem to keep free-tier version a viable or only option to many end-users, particularly those in the low-income countries. However, despite the wide use of free-tier versions, the paid versions usually have better performance or less variation in performance. Second, our analyses include randomly selected 50 original articles from five environmental science journals from 2024 to 2025. While year seems not linked to the retrieval errors/performance, our findings should be considered preliminary due to the small sample size. Caution should also be used when interpreting and generalizing our findings. Large and diverse studies are needed to examine our findings. Third, the relevance score is overall objective, but relies on the original article's reference list. Thus, there is a major limitation of reduced sensitivity since some relevant articles retrieved by LLM may not be cited. On the other hand, human interpretation seems too subjective to us despite its higher sensitivity. It remains challenging and unclear how to best balance the objectivity and sensitivity. Future methodological studies are needed to address this important question. Finally, although the validation process was structured and overall objective, some fields may be subjective and have unintended biases.

### 4.3 Future directions

It will be interesting to assess the effects of these erroneous references on manuscript review, citations, and research consensus among scientists. However, the bigger concern to us is that these erroneous references may be incorporated into our research community and bibliography. While careful authors and reviewers may be able to recognize and remove them, journal editors should still consider and eliminate errors in LLM-assisted literature retrieval.

It is also interesting and important to understand whether the similar advantages of abstract input-method are present in other journals and fields. Maybe, environmental science is less sampled and trained on by the LLM platform and thus has different retrieval validity of abstract versus full-text input.

Future system may improve reference retrieval by combining LLMs with specific retrieval databases. For example, RAG-based systems have shown that external retrieval and citation-grounded generation can improve scientific synthesis accuracy [1, 3, 9]. Recently developed RAG with adaptive routing has more favorable effectiveness-efficiency ratios than the fixed paradigm approach [35], and may also help with LLM-assisted literature retrieval. However, future studies should test whether these systems improve reference retrieval accuracy in environmental science. They should also evaluate whether automated verification of DOI, EID, and bibliographic information can reduce retrieval errors while our prompt of verifying the retrieved bibliographic information did not.

## 5. Conclusions

LLMs may assist with literature exploration, but their performance in literature retrieval of environmental science remains moderate and inconsistent across journal and LLM platform. Abstract based prompt-input seems to be associated with higher score ratios and fewer errors. It thus may be preferred over full-text one, while caution must be used due to small sample size of this study. While the retrieved reference on top of the output list more likely have accurate and relevant to the index article, verification of LLM-retrieved references still seems necessary. Additional and large studies are needed to examine our findings in environmental science and other fields.

# Author Contributions

YC: data curation, visualization, formal analysis, writing—original draft preparation; YZ: conceptualization, methodology, validation; LZ: formal analysis, conceptualization, methodology, supervision, validation; All authors: writing—review and editing. All authors have read and agreed to the submitted version of the manuscript.

# Funding

This research received no external funding.

# Institutional Review Board Statement

Not applicable. This study did not involve human participants, human tissue, human data requiring consent, or animals.

# Informed Consent Statement

Not applicable.

# Data Availability Statement

The cleaned data, analysis code, and generated figures used in this study are available from the corresponding author upon reasonable request and will be deposited in [repository name/link] before final publication if required by the journal.

# Conflicts of Interest

The authors declare no conflict of interest.

**Table 1. Performance of LLM-assisted literature retrieval in environmental science by inputting method.**

| | | Abstract | | Full text | | P value |
|---|---|---|---|---|---|---|
| Metrics | | Count | % | Count | % | |
| DOI score | | | | | | |
| | 0 | 1164 | 38.8 | 1343 | 44.77 | <0.001 |
| | 1 | 1836 | 61.2 | 1657 | 55.23 | |
| EID score | | | | | | |
| | 0 | 2538 | 84.6 | 2726 | 2538 | <0.001 |
| | 1 | 462 | 15.4 | 274 | 462 | |
| Google Scholar score | | | | | | |
| | 0 | 1712 | 57.07 | 1751 | 58.37 | 0.308 |
| | 1 | 1288 | 42.93 | 1249 | 41.63 | |
| Relevance score | | | | | | |
| | 0 | 2257 | 75.23 | 2214 | 73.8 | 0.203 |
| | 1 | 743 | 24.77 | 786 | 26.2 | |
| Complete fabrication | | | | | | |
| | 0 | 2166 | 72.2 | 2088 | 69.6 | 0.027 |
| | 1 | 834 | 27.8 | 912 | 30.4 | |
| Score ratio* | | 0.43 | 0.32 | 0.40 | 0.32 | <0.001 |

DOI: Digital object identifier; EID, **Scopus Electronic Identifier.**

*The row of score ratio shows mean with standard deviation.

**Table 2. Multilevel mixed-effects multivariable regression on the factors associated with score ratio.**

| | coefficient | 95% CI | P |
|---|---|---|---|
| **LLM Platform** | | | |
| Chat GPT | Reference | | |
| Claude | 0.22 | (0.17 - 0.27) | <0.001 |
| Deepseek | -0.03 | (-0.08 - 0.03) | 0.322 |
| Gemini | -0.28 | (-0.34 - -0.23) | <0.001 |
| Grok | 0.18 | (0.13 - 0.24) | <0.001 |
| Perplexity | -0.1 | (-0.15 - -0.05) | <0.001 |
| **Input method** | | | |
| Abstract | Reference | | |
| Full text | -0.03 | (-0.04 - -0.02) | <0.001 |
| **Journal** | | | |
| *Energy and Environmental Science* | Reference | | |
| *Environmental Science and Technology* | 0.04 | (-0.01 - 0.09) | 0.081 |
| *Nature Climate Change* | 0.02 | (-0.03 - 0.07) | 0.452 |
| *Nature Sustainability* | 0.05 | (-0.00 - 0.10) | 0.057 |
| *Lancet Planetary Health* | 0.02 | (-0.03 - 0.07) | 0.369 |
| **Output sequence^** | -0.03 | (-0.03 - -0.03) | <0.001 |

LLM, large language model; CI, confidence interval of the odds ratio.

^ Output sequence was used as a continuous data, while others as categorical data.

**Table 3. Multilevel mixed-effects multivariable logistic regression on the factors associated with complete fabrication.**

| | Odds ratio | 95% CI | P |
|---|---|---|---|
| **LLM Platform** | | | |
| Chat GPT | Reference | | |
| <0.001 | <0.001 | <0.001 | <0.001 |
| Deepseek | 0.74 | (0.60 - 0.91) | 0.004 |
| Gemini | 7.42 | (6.04 - 9.11) | <0.001 |
| Grok | 0.19 | (0.15 - 0.25) | <0.001 |
| Perplexity | 1.78 | (1.46 - 2.16) | <0.001 |
| **Input method** | | | |
| Abstract | Reference | | |
| Full text | 1.20 | (1.05 - 1.37) | 0.008 |
| **Journal** | | | |
| *Energy and Environmental Science* | Reference | | |
| *Environmental Science and Technology* | 0.39 | (0.32 - 0.49) | <0.001 |
| *Nature Climate Change* | 0.76 | (0.62 - 0.93) | 0.008 |
| *Nature Sustainability* | 0.74 | (0.61 - 0.91) | 0.004 |
| *Lancet Planetary Health* | 0.41 | (0.33 - 0.51) | <0.001 |
| **Output sequence^** | 1.22 | (1.19 - 1.25) | <0.001 |

LLM, large language model; CI, confidence interval of the odds ratio.

^ Output sequence was used as a continuous data, while others as categorical data.